\documentclass[twocolumn,showpacs,preprintnumbers,amsmath,amssymb]{revtex4}
\usepackage{graphicx}% Include figure files
\usepackage{dcolumn}% Align table columns on decimal point
\usepackage{bm}% bold math
\usepackage{psfrag}
\newcommand{\eref}[1]{(\ref{#1})}

\newcommand{\rmi}{\mathrm i}
\newcommand{\rme}{\mathrm e}
\newcommand{\e}{\rme}
\newcommand{\rmd}{\mathrm d}

\renewcommand{\Im}{\mathrm{Im}}
\renewcommand{\vec}{\boldsymbol}

\newcommand{\op}[1]{\overline{\vec{#1}}}
\newcommand{\optilde}[1]{\widetilde{\vec{#1}}}

\newcommand{\ep}{\epsilon}

\newcommand{\hvec}[1]{\hat{\vec{#1}}}

\newcommand{\intr}[1]{\int  \rmd^3 \rr_{#1}\,}
\newcommand{\iintr}[2]{\int  \rmd^3 \rr_{#1}\, \rmd^3 \rr_{#2}\,}

\newcommand{\intk}[1]{\int \frac{\rmd^3 \vec{k}_{#1}}{(2\pi)^3}\,}

\newcommand{\no}{\nonumber}
\newcommand{\rr}{\vec{r}}

\newcommand{\E}{\vec{E}}

\newcommand{\kk}{\vec{k}}

\begin{document}

\preprint{}

\title{Effective dielectric constant for a random medium}

\author{A. Soubret}
\affiliation{NOAA, Environmental Technology Laboratory, 325 Broadway, Boulder CO 80305-3328.}%Lines break automatically or can be forced with \\
\email{asoubret@hms.harvard.edu}
\author{G. Berginc}%
\affiliation{Thal\`es Optronique, Bo\^{\i}te Postale 55, 78233 Guyancourt Cedex,France
}%

\date{\today}% It is always \today, today,
             %  but any date may be explicitly specified

\begin{abstract}
In this paper, we present an approximate expression for
determining the effective permittivity describing the coherent
propagation of  an electromagnetic wave in  random media. Under
the Quasicrystalline Coherent Potential Approximation (QC-CPA), it
is known that multiple scattering theory provided an expression
for this effective permittivity. The numerical evaluation of this
one is, however, a challenging problem. To find a tractable
expression, we add some new approximations to the (QC-CPA)
approach. As a result, we obtained an expression for the effective
permittivity which contained at the same time the Maxwell-Garnett
formula in the low frequency limit, and the Keller formula, which
has been recently proved to be in good agreement for particles
exceeding the wavelength.
\end{abstract}

\pacs{42.25.Bs, 41.20.Jb, 78.20.-e}% PACS, the Physics and Astronomy
                             % Classification Scheme.
%\keywords{Suggested keywords}%Use showkeys class option if keyword
                              %display desired
\maketitle
\section{\label{Intro} Introduction}
The description  of electromagnetic waves propagation in random
media in term of the properties of the constituents has been
studied extensively in the past decades
~\cite{Chandra,Hulst1,Hulst2,Bohren,Ishi2,Frish,Lag1,Apresyan,Rytov4,Bara1,Sheng1,Sheng2,Kong,%
Kong2001-1,Kong2001-2,Kong2001-3,Ulaby3,Fung,%
Garland,AMS,Berthier,Scaife,Choy,Sihvola}. In most of works, the
basic idea is to calculate several statistical moments of the
electromagnetic field to understand how the wave interact with the
the random medium~\cite{Apresyan,Sheng1,Sheng2,Kong,Kong2001-3}.
In this paper, we are concerned by the first moment which is the
average electric field. Under some assumption, it can be shown
that the average electric field propagates as if the  medium where
homogeneous but with a renormalized permittivity, termed
\emph{effective permittivity}. The calculation of this parameter
as a long history which dates back from the work of
Clausius-Mossotti and Maxwell Garnett~\cite{Sihvola}. Since then,
most of the study are concerned with the quasi-static limit where
retardation effect are
neglected~\cite{Berthier,Scaife,Sihvola,Choy,Liebsch,Cum,Stroud1,Stroud2,Felderhof,Barrera,Agarwal1,%
Lamb,Mochan1,Mochan2}. In order to take into account scattering
effects, quantum multiple scattering theory has been transposed in
the electromagnetic
case~\cite{Frish,Apresyan,Sheng1,Sheng2,Kong,Kong2001-3}, but as a
rigorous analytical answer is unreachable, several approximation
schemes have been
developed~\cite{Frish,Apresyan,Sheng2,Tsang6,Kong,Kong2001-3,Lax,Waterman,Korringa,%
Gyorffy,Soven,Davis,Golden,Jing,Bush,Walden1,Walden2,Stoyanov1,Stoyanov2}.
One of the most advanced is the Quasicrystalline Coherent
Potential Approximation (QC-CPA) which takes into account the
correlation between the
particles~\cite{Tsang6,Kong,Kong2001-3,Lax,Waterman,Korringa,Gyorffy,Soven}.
Unfortunately, except at low frequency, the answer is still too
involved to permit the calculation of the effective permittivity.

The aim of this paper is to add some new approximations to the
(QC-CPA) approach which furnish a tractable equation for the
effective permittivity. The expression obtained contains the low
frequency limit of the (QC-CPA) approach. At this limit, the
(QC-CPA) equations can be written as a generalized Maxwell Garnett
formula and are proven to be in good agreement with the
experimental
results~\cite{Mandt2,Zurk,West2,Kong2001-2,Kong2001-3}.
Furthermore, the formula obtained contains also the approximate
formula due to Keller, which has been derived in using scalar
theory, but seems to  be in accord with the experimental data for
particles larger than a wavelength~\cite{Keller1,Hespel,Hespel2}.

The paper is organized as follows. In Section \ref{SecDyson}, we
introduce the multiple scattering formalism and we show under what
hypothesis the effective medium theory is valid. In section
\ref{secCPA}, we recall the different steps in order to obtain the
system of equation verified by the effective permittivity under
the (QC-CPA) approach. Then, we introduce, in section
\ref{Newapprox}, some new approximations in order to obtain a
tractable formula for the effective permittivity. In the two
following section \ref{Rayleigh}, \ref{Keller}, we derive
respectively the low frequency and high frequency limit of our new
approach.

\section{Dyson equation and effective permittivity}
\label{SecDyson} In the following, we consider harmonic waves with
$\e^{-\rmi\omega t}$ pulsation. We consider an ensemble of $N\gg
1$ identical spheres of radius $r_s$ with dielectric function
$\ep_s(\omega)$ within a infinite medium with dielectric function
$\ep_1(\omega)$. The field produced at $\rr$ by a discrete source
located at $\rr_0$ is given by the dyadic Green function
$\op{G}(\rr,\rr_0,\omega)$, which verifies the following
propagation equation:
\begin{align}
\nabla\times\nabla\times
\op{G}(\rr,\rr_0,\omega)-\ep_V(\rr,\omega)\,K_{vac}^2\,\op{G}(\rr,\rr_0,\omega)\no\\
=\delta(\rr-\rr_0)\op{I}\label{defGrr0}
\end{align}
where $K_{vac}=\omega/c$ with $c$ the speed of light in vaccum and
\begin{displaymath}
\ep_V(\rr,\omega)=\ep_1(\omega)+\sum_{j=1}^N
[\ep_s(\omega)-\ep_1(\omega)]\,\Theta_s(\rr-\rr_j)\,,\label{eVrr}
\end{displaymath}
where $\rr_1,\dots,\rr_N$ are the center of the particles and
$\Theta_s$ describes the spherical particle shape :
\begin{equation}
\Theta_s(\rr)=\left\{\begin{array}{cc}1 & \mbox{if}\quad ||\rr||<r_s\\
  0 & \mbox{if}\quad ||\rr||>r_s\end{array}\right.\,.\label{Thetad}
\end{equation}
The solution of equation \eref{defGrr0} is uniquely defined if
we impose the radiation condition at infinity.

The multiple scattering process by the particles is mathematically
decomposed in introducing the Green function $\op{G}_1^{\infty}$,
describing the propagation within  an homogenous medium with
permittivity $\ep_1(\omega)$, which verifies the following
equation:
\begin{align}
\nabla\times\nabla\times
\op{G}_1^{\infty}(\rr,\rr_0,\omega)-\ep_1(\omega)\,K_{vac}^2\,\op{G}_1^{\infty}(\rr,\rr_0,\omega)\,\no\\
=\delta(\rr-\rr_0)\op{I}\,,
\end{align}
with the appropriate   boundary conditions. In an infinite random
medium, we have~\cite{Bladel,Tai,Kong2001-3}:
\begin{equation}
\op{G}^{\infty}_1(\rr,\rr_0,\omega)=\left[\op{I}+\frac{\nabla\nabla}{K_{1}^2}\right]
\frac{e^{\rmi\,K_{1}\,||\rr-\rr_0||}}{4\pi||\rr||}
\end{equation}
where  $K_1^2=\ep_1(\omega)\,K_{vac}^2$.

In using this Green function, we decompose the Green function
$\op{G}(\rr,\rr_0,\omega)$ under the following
form~\cite{Frish,Kong,Kong2001-3,Sheng1}:
\begin{align}
\op{G}=\op{G}_1^{\infty}+\op{G}_1^{\infty}\cdot\op{V}\cdot\op{G}\,,\label{G1VG}
\end{align}
where the following operator notation is used:
\begin{equation}
[\op{A}\cdot\op{B}](\rr,\rr_0)=\int
\rmd^3\,\rr_1\,\op{A}(\rr,\rr_1)\cdot\op{B}(\rr_1,\rr_0)\,.\label{Chap3Produitr}
\end{equation}
The potential $\op{V}$, which describes the interaction between
the wave and the particles, is given by:
\begin{align}
\op{V}&=\sum_{i=1}^N\op{v}_{\rr_i}\,,\\
\op{v}_{\rr_i}(\rr,\rr_0,\omega)&=(2\pi)^2\,\delta(\rr-\rr_0)\,\op{v}_{\rr_i}(\rr,\omega)\,,\\
\op{v}_{\rr_i}(\rr,\omega)&=[K^2_s-K^2_1]\,\Theta_d(\rr-\rr_i)\op{I}\,.
\end{align}
with  $K_s^2=\ep_s(\omega)\,K_{vac}^2$. It is useful to introduce
the T matrix defined by~\cite{Frish,Kong,Kong2001-3,Sheng1}:
\begin{align}
\op{G}=\op{G}_1^{\infty}
+\op{G}_1^{\infty}\cdot\op{T}\cdot\op{G}^{\infty}_1\,.\label{Deftmatrix}
\end{align}
In iterating equation \eref{G1VG} and comparing it with the
definition \eref{Deftmatrix}, we show that the T matrix verifies
the following equation:
\begin{equation}
\op{T}=\op{V}+\op{V}\cdot\op{G}_1^{\infty}\cdot\op{T}\,.\label{VG1T}
\end{equation}
If we introduce the T matrix of each scatterer by:
\begin{eqnarray}
\op{t}_{\rr_i}=\op{v}_{\rr_i}+\op{v}_{\rr_i}\cdot\op{G}_{1}^{\infty}\cdot\op{t}_{\rr_i}\,,\\
\end{eqnarray}
we can decompose the T matrix for the whole system, in a series of
multiple scattering processes by the
particles~\cite{Sheng1,Kong,Kong2001-3,Frish}:
\begin{eqnarray}
\op{T}=\sum_{i=1}^N\op{t}_{\rr_i}+\sum_{i=1}^N\sum_{j=1,j\neq i}^N
\op{t}_{\rr_j}\cdot\op{G}_1^{\infty}\cdot\op{t}_{\rr_i}+\cdots\,.\label{DevT}
\end{eqnarray}
This T matrix is useful to calculate the average field $<\op{G}>$
since we have:
\begin{align}
<\op{G}>=\op{G}_1^{\infty}
+\op{G}_1^{\infty}\cdot<\op{T}>\cdot\op{G}^{\infty}_1
\end{align}
The equivalent of the potential operator $\op{V}$ for the average
Green function $<\op{G}>$ is the mass operator $\op{\Sigma}$
defined by:
\begin{align}
<\op{G}>=\op{G}_1^{\infty}
+\op{G}_1^{\infty}\cdot\op{\Sigma}\cdot<\op{G}>\,.\label{G1SigG}
\end{align}
Similarly to equation \eref{VG1T}, we have the following
relationship between the average T matrix and the mass operator:
\begin{equation}
<\op{T}>=\op{\Sigma}+\op{\Sigma}\cdot\op{G}_1^{\infty}\cdot<\op{T}>
\end{equation}
or equivalently,
\begin{equation}
\op{\Sigma}=<\op{T}>\cdot\left[\op{I}+\op{G}_1^{\infty}\cdot<\op{T}>\right]^{-1}\,.\label{defSigma}
\end{equation}
The mass operator correspond to all irreducible diagrams in the
Feynman representation~\cite{Frish,Apresyan,Kong,Sheng1}. The
equation \eref{G1SigG} written in differential form is:
\begin{align}
\nabla\times\nabla\times
<\op{G}(\rr,\rr_0,\omega)>-\ep_1(\omega)\,K_{vac}^2\,<\op{G}(\rr,\rr_0,\omega)>\no\\
-\intr{1}\op{\Sigma}(\rr,\rr_1,\omega)\cdot<\op{G}(\rr_1,\rr_0,\omega)>=\delta(\rr-\rr_0)
\op{I}\,.\label{Dyson}
\end{align}
For a statistical homogeneous medium we have:
\begin{align}
\op{\Sigma}(\rr,\rr_1,\omega)&=\op{\Sigma}(\rr-\rr_1,\omega)\,,\\
<\op{G}(\rr,\rr_0,\omega)> &=<\op{G}(\rr-\rr_0,\omega)>\,.
\end{align}
Thus, we can use a Fourier transform:
\begin{eqnarray}
\op{\Sigma}(\kk,\omega)=\intr{}\exp(-\rmi\kk\cdot\rr)\,\op{\Sigma}(\rr,\omega)\,,\\
\op{G}(\kk,\omega)=\intr\,\exp(-\rmi\,\kk\cdot\rr)\,\op{G}(\rr,\omega)\,,
\end{eqnarray}
and equation \eref{Dyson} becomes:
\begin{equation}
\left[||\vec{k}||^2\,(\op{I}-\hvec{k}\hvec{k})-\ep_1(\omega)\,K_{vac}^2\op{I}-
\op{\Sigma}(\vec{k},\omega)\right]\cdot<\op{G}(\vec{k},\omega)>=\op{I}\,.\label{Dyson1}
\end{equation}
For a statistical isotropic medium, we have:
\begin{equation}
\op{\Sigma}(\kk,\omega)=\Sigma_{\perp}(||\kk||,\omega)(\op{I}-\hvec{k}\hvec{k})+
\Sigma_{\parallel}(||\kk||,\omega)
\,\hvec{k}\hvec{k}\,.\label{Propiso}
\end{equation}
with $\hvec{k}=\vec{k}/||\vec{k}||$ and then:
\begin{align}
<\op{G}(\vec{k},\omega)>&=\left[||\vec{k}||^2\,(\op{I}-\hvec{k}\hvec{k})-\ep_1(\omega)\,K_{vac}^2\op{I}-
\op{\Sigma}(\vec{k},\omega)\right]^{-1}\,,\\
&=\frac{\op{I}-\hvec{k}\hvec{k}}{||\vec{k}||^2\no
-(\ep_1(\omega)K_{vac}^2+\Sigma_\perp(||\vec{k}||,\omega))}\no\\
&-\frac{\hvec{k}\hvec{k}}{\ep_1(\omega)K_{vac}^2+\Sigma_\parallel(||\vec{k}||,\omega)}\label{eqDys1}
\end{align}

In the following, %we will suppose that
%$\Sigma_{\perp}(||\kk||,\omega)=\Sigma_{\parallel}(\kk,\omega)$,
we introduce two effective permittivity function $\ep^{\perp}_e$
and $\ep_{e}^{\parallel}$ defined by:
\begin{align}
\ep^{\perp}_e({\scriptstyle
||\vec{k}||},\omega)K_{vac}^2 &=\ep_1(\omega)K_{vac}^2+\Sigma_\perp(||\vec{k}||,\omega)\,,\\
\ep^{\parallel}_e({\scriptstyle ||\vec{k}||},\omega)K_{vac}^2&
=\ep_1(\omega)K_{vac}^2+\Sigma_\parallel(||\vec{k}||,\omega)\,,
\end{align}
 and \eref{eqDys1} is written:
\begin{align}
&<\op{G}(\vec{k},\omega)>\no\\&=\left[\op{I}-\frac{\vec{k}\vec{k}}{\ep^\perp_e({
\scriptstyle ||\vec{k}||},\omega)K_{vac}^2}\right]
\frac{1}{||\vec{k}||^2-\ep_e^\perp({\scriptstyle
||\vec{k}||},\omega)K_{vac}^2}\no\\
&+\frac{\hvec{k}\hvec{k}}{\ep_e^\perp
(||\vec{k}||,\omega)K_{vac}^2}
-\frac{\hvec{k}\hvec{k}}{\ep_e^\parallel(||\vec{k}||,\omega)K_{vac}^2}\,.
\label{eqDys2}
\end{align}
The Green function in the space domain is:
\begin{align}
&<\op{G}(\rr,\omega)>\no\\
&=\intk{}\left[\op{I}+\frac{\nabla\nabla}{\ep_e^\perp({\scriptstyle
||\vec{k}||},\omega)K_{vac}^2}\right]
\frac{\e^{\rmi\,\vec{k}\cdot\rr}}{||\vec{k}||^2-\ep_e^\perp({\scriptstyle
||\vec{k}||},\omega)K_{vac}^2}\no\\
&+\intk{}\left[\frac{\e^{\rmi\,\vec{k}\cdot\rr}}{\ep_e^\perp
(||\vec{k}||,\omega)K_{vac}^2}-\frac{\e^{\rmi\,\vec{k}\cdot\rr}}{\ep_e^{\parallel}(||\vec{k}||,\omega)K_{vac}^2
}\right]\,\frac{\vec{k}\vec{k}}{||\vec{k}||^2}\,.\label{gromega}
\end{align}
After an integration on the solid angle in equation
\eref{gromega} given by the expression \eref{App1} in the
appendix, we obtain:
\begin{align}
&<\op{G}(\rr,\omega)>\no\\
&=\frac{1}{\rmi||\rr||}\int_{-\infty}^{+\infty} \frac{\rmd
K}{(2\pi)^2}\left[\op{I}+\frac{\nabla\nabla}{\ep_e^\perp(K,\omega)K_{vac}^2}\right]
\frac{K\,\e^{\rmi\,K\,||\rr||}}{K^2-\ep_e^\perp(K,\omega)K_{vac}^2}\no\\
&+\frac{1}{\rmi||\rr||}\nabla\nabla\int_{-\infty}^{+\infty}
\frac{\rmd
K}{(2\pi)^2}\left[\frac{\e^{\rmi\,K\,||\rr||}}{\ep_e^\perp(K,\omega)K_{vac}^2}-
\frac{\e^{\rmi\,K\,||\rr||}}{\ep_e^{\parallel}(K,\omega)K_{vac}^2}\right]\frac{1}{K}\,\label{eqIntKG}
\end{align}
where we have supposed that $\ep_e^\perp({\scriptstyle
||\vec{k}||},\omega)=\ep_e^\perp({\scriptstyle
-||\vec{k}||},\omega)$ and $\ep_e^\parallel({\scriptstyle
||\vec{k}||},\omega)=\ep_e^\parallel({\scriptstyle
-||\vec{k}||},\omega)$ . In using the residue theorem, we  easily
evaluate these integrals. However, we neglect the longitudinal
excitation, which are solutions of $\ep_e^\perp(K,\omega)=0$ and
$\ep_e^\parallel(K,\omega)=0$, since we are only interested by the
propagation of the transversal field. Furthermore, we see that the
contribution due to pole $K=0$ in the second term of equation
\eref{eqIntKG} is null; in fact, the dyadic $\nabla\nabla$ operate
on a constant since we have $\e^{\rmi\,K\,||\rr||}=1$ for this
pole. Hence, we obtain the following expression for the Green
function:
\begin{equation}
<\op{G}(\rr,\omega)>=\sum_{i=1}^{n}\left[\op{I}+
\frac{\nabla\nabla}{K_{e\,i}^2}\right]\frac{e^{\rmi\,K_{e\,i}\,||\rr||}}{4\pi||\rr||}\,,
\label{DevpoleG}
\end{equation}
where  $K_{e\,i}$ are  the roots of
$K_{e\,i}^2=\ep_e^\perp(K_{e\,i},\omega)K^2_{vac}$ such as
$\Im(K_{e\,i})>0$ to insure that the radiation condition at
infinity is verified. Sheng has called the roots $K_{e\,i}$  the
quasi-modes of the random media~\cite{Sheng1,Jing}. If we only
consider the root $K_e=K_{e\,j}$ which has the smallest imaginary
part ($\Im(K_{e\,j})=min_i\left[\Im(K_{e\,i})\right]$) and then
the smallest exponential  factor in equation \eref{DevpoleG},
we define the effective permittivity by
$\ep_e(\omega)=\ep_e^\perp(K_e,\omega)$. The average Green
function is then equal to the Green function for an infinite
homogenous medium with permittivity $\ep_e(\omega)$ :
\begin{equation}
<\op{G}(\rr,\omega)>=\op{G}^\infty_e(\rr,\omega)\,,
\end{equation}
where
\begin{equation}
\op{G}^\infty_e(\rr,\omega)=\left[\op{I}+\frac{\nabla\nabla}{K_{e}^2}\right]
\frac{e^{\rmi\,K_{e}\,||\rr||}}{4\pi||\rr||}\,.
\end{equation}
Thus, the effective medium approach is valid if we neglect the
longitudinal excitation in the medium and if the propagative mode
with the smallest imaginary part is the primary contribution in
the developpement \eref{DevpoleG}.
\section{The Coherent-Potential and Quasi-Crystalline Approximations}
\label{secCPA} Previously, we have shown how the mass operator is
related to the effective permittivity.  To calculate the mass
operator, we can use   equations \eref{DevT} and
\eref{defSigma}. However, we can improve this system of equations
in rewriting the Green function development \eref{G1VG} in
replacing the Green function $\op{G}_1^{\infty}$  by
$\op{G}_e^\infty$:
\begin{align}
\op{G}=\op{G}_e^{\infty}+\op{G}_e^{\infty}\cdot\op{V}_e\cdot\op{G}\,,
\end{align}
where we have to introduce a new potential $\op{V}_e$:
\begin{align}
\op{V}_e&=\sum_{i=1}^N\op{v}_{e,\rr_i}\,,\\
\optilde{v}_{e,\rr_i}(\rr,\rr_0,\omega)&=(2\pi)^2\,\delta(\rr-\rr_0)\,\optilde{v}_{e,\rr_i}(\rr,\omega)\,,\\
\optilde{v}_{e,\rr_i}(\rr,\omega)&=[K^2_s-K^2_e]\,\Theta_s(\rr-\rr_i)\op{I}\no\\
&+[K^2_1-K^2_e]\op{I}\,.\label{defvtilde_e}
\end{align}
Similarly to the previous section, we introduce a T matrix such
that:
\begin{align}
\op{G}=\op{G}_e^{\infty}
+\op{G}_e^{\infty}\cdot\op{T}_e\cdot\op{G}^{\infty}_e\,.\label{DefTe}
\end{align}
which admits the following decomposition:
\begin{equation}
\op{T}_e=\sum_{i=1}^N\optilde{t}_{e;\rr_i}+\sum_{i=1}^N\sum_{j=1,j\neq
i}^N
\optilde{t}_{e;\rr_j}\cdot\op{G}_e^{\infty}\cdot\optilde{t}_{e;\rr_i}+\cdots\label{DevTe}
\end{equation}
where we have defined a renormalized T matrix for the particles:
\begin{eqnarray}
\optilde{t}_{e,\rr_i}=\optilde{v}_{e,\rr_i}+\optilde{v}_{e,\rr_i}\cdot\op{G}_{e}^{\infty}\cdot
\optilde{t}_{e,\rr_i}\,.\label{devte}
\end{eqnarray}
In supposing that the effective medium approach is correct, we
impose the following condition on the average field:
\begin{equation}
<\op{G}(\rr,\omega)>=\op{G}_e(\rr,\omega)\,,\label{CondCPA}
\end{equation}
or equivalently,
\begin{equation}
<\op{T}_e>=\op{0}\,,\label{PropTe}
\end{equation}
 due to equation \eref{DefTe}.
 The condition \eref{CondCPA} is the Coherent-Potential
 Approximation (CPA)~\cite{Soven,Tsang6,Sheng2,Kong,Kong2001-3}.
The expression \eref{PropTe} and \eref{DevTe} form a closed system
of equations on the unknown permittivity $\ep_e(\omega)$. To the
first order in density of particles, this system of equations
gives equation:
\begin{equation}
\sum_{i=1}^N<\optilde{t}_{e,\rr_i}>=\vec{0}\,.\label{condCPA-1}
\end{equation}
In Fourier-space, the T matrix for one scatterer verifies the
property:
\begin{equation}
\optilde{t}_{e,\rr_i}(\vec{k}|\vec{k}_0)=\e^{-\rmi(\vec{k}-\vec{k}_0)\cdot\rr_i}
\,\optilde{t}_{e,o}(\vec{k}|\vec{k}_0)\,,\label{ttranslation}
\end{equation}
where  $\optilde{t}_{e,o}$ is the T matrix for a particle located
at the origin of coordinate. The average of the exponential term,
introduced in  equation \eref{condCPA-1} by the properties
\eref{ttranslation}, gives for a statistical homogeneous medium:
\begin{align}
<\sum_{i=1}^N \e^{-\rmi(\vec{k}-\vec{k}_0)\cdot\rr_i}>&=N
\intr{}\frac{1}{\mathcal{V}}
\e^{-\rmi(\vec{k}-\vec{k}_0)\cdot\rr}\,,\\
&=n \,(2\pi)^2\,\delta(\vec{k}-\vec{k}_0)
\end{align}
where we have introduced the density of scatterers
$n=N/\mathcal{V}$ with $\mathcal{V}$ the volume of the random
medium. The condition \eref{condCPA-1} becomes:
\begin{equation}
\optilde{t}_{e,o}(\vec{k}_0|\vec{k}_0)=\op{0}\,.\label{CondCPASheng}
\end{equation}
This CPA condition has been used in several
works~\cite{Jing,Sheng1,Soukoulis2}. It is worth mentioning that
operator $\optilde {t}_{e,\rr_i}(\vec{k}|\vec{k}_0)$ is not the T
matrix describing the scattering by a particle of permittivity
$\ep_s(\omega)$ surrounded by a medium of permittivity
$\ep_e(\omega)$. To do so, the operator \eref{defvtilde_e} should
have the following form:
\begin{align}
\optilde{v}_{e,\rr_i}(\rr,\omega)=[K^2_s(\omega)-K^2_e(\omega)]\,\Theta_s(\rr-\rr_i)\op{I}\,.
\label{defVe-2}
\end{align}
However, we see that the operator
$\optilde{v}_{e,\rr_i}(\rr,\omega)$ is quiet different from the
equation \eref{defvtilde_e}, and in particular we have
 $\optilde{v}_{e;\rr_i}(\rr,\omega)=[K^2_1-K^2_e]\op{I}$ for
 $||\rr-\rr_i||>r_s$ contrary to the definition \eref{defVe-2},
 where $\optilde{v}_{e;\rr_i}(\rr,\omega)=\vec{0}$ when $\rr$ is
outside the particle. Thus, the operator $\optilde{t}_{e,\rr_i}$
is non-local and cannot be obtained from the classical Mie
theory~\cite{Hulst1,Bohren,Tsang6}. To overcome this difficulty,
 the
operator $\optilde{t}_{e,\rr_i}$ is replaced by the scattering
operator of a "structural unit" in the
works~\cite{Sheng1,Soukoulis2,Berthier}. Nevertheless, this
approach doesn't seem to have any theoretical justification.

Hence, we prefer to use the more rigorous approach introduce in
the scattering theory by disorder liquid
metal~\cite{Gyorffy,Korringa} and adapted in the electromagnetic
case by Tsang \emph{et al.}~\cite{Kong,Kong2001-3}. In this
approach, the non-local term $[K^2_1-K^2_e]\op{I}$ is correctly
taking into account by averaging equations \eref{DevTe} where
the correct potential $\optilde{v}_{e,\rr_i}$, defined by
\eref{defvtilde_e}, is used. A system of hierarchic equations is
obtained where correlation functions between two or more particles
are successively introduced. The chain of equations is closed in
using the Quasi-Crystalline Approximation (QCA), which neglect the
fluctuation of the effective field, acting on a particle located
at $\rr_j$, due to a deviation of a particle located at $\rr_i$
from its average position~\cite{Lax}. This approximation describes
the correlation between the particles, only with a two-point
correlation function $g(\rr_i,\rr_j)=g(||\rr_i-\rr_j||)$. Under
the QC-CPA scheme, we obtain the following expression for the mass
operator~\cite{Korringa,Gyorffy,Tsang6,Kong,Kong2001-3}:
\begin{align}
&\op{\Sigma}(\vec{k}_0,\omega)=n\,\op{C}_{e,o}(\kk_0|\kk_0)\,,\label{PropSigma}
\\ &\op{C}_{e,o}(\kk|\kk_0)=\op{t}_{e,o}(\kk|\kk_0)\no\\
&+n\,\intk{1}\,h(\kk-\kk_1)\,\op{t}_{e,o}(\kk|\kk_1)\cdot\op{G}_{e}^{\infty}(\kk_1)
\cdot\op{C}_{e,o}(\kk_1|\kk_0)\,
\label{expCkk0}
\end{align}
where
\begin{align}
\op{t}_{e,o}&=\op{v}_{e,o}+\op{v}_{e,o}\cdot\op{G}_{e}^{\infty}\cdot\op{t}_{e,o}\,,\\
\op{v}_{e,o}(\rr,\rr_0)&=(2\,\pi)\delta(\rr-\rr_0)\,\op{v}_{e,o}(\rr)\,,\\
\op{v}_{e,o}(\rr)&=[K^2_s-K^2_1]\,\Theta_s(\rr)\op{I}\label{Defve-2}
\end{align}
 and
\begin{align}
h(\rr)&=g(\rr)-1\,,\\
h(\kk-\kk_1)&=\intr\,\exp(-\rmi\,(\kk-\kk_1)\cdot\rr)\,h(\rr)\,,\\
\op{G}_{e}^{\infty}(\kk)&=\intr{}\exp(-\rmi\kk\cdot\rr)\,\op{G}_{e}^{\infty}(\rr)\,.\label{fourierG1}
\end{align}
If we rewrite the potential  \eref{Defve-2} under the following
form:
\begin{equation}
\op{v}_{e;o}(\rr)=[\tilde{K}^2_s-K^2_e]\,\Theta_s(\rr)\op{I}\label{Defve-3}
\end{equation}
where we have defined a new wave number $\tilde
K^2_s=K_s^2-K^2_1+K^2_e$, we see that the operator $\op{t}_{e,o}$
is the T matrix for a scatterer of permittivity $\tilde
\ep_s=\ep_s-\ep_1+\ep_e$ in a medium of permittivity $\ep_e$.

As it is described in the previous section, the effective
propagation $K_e$ constant is the root,which has the smallest
imaginary part, of the  equation:
\begin{equation}
K_e^2=K_1^2+\Sigma_\perp(K_e,\omega)\,,\label{EqKe}
\end{equation}
where the mass operator is decomposed under the form
\eref{Propiso}. Once the effective wave number $K_e$ obtained, the
effective permittivity is given by:
\begin{equation}
\ep_e(\omega)=K_e^2/K_{vac}^2\,.\label{Obtentionepe}
\end{equation}
\section{Some further approximations}
\label{Newapprox} As it can be guessed, solving numerically the
previous system of equations (\ref{PropSigma}-\ref{Obtentionepe})
is full of complexities. However, the low frequency limit of this
system of
 equation
has been obtained analytically and has  shown to be in good
agreement with the experimental results~\cite{Kong,Kong2001-3}. We
have also to mention that  the numerical solution of the
quasicrystalline approximation (but without the coherent potential
approximation) has been developed~\cite{Kong2001-2,Zurk}.

\begin{figure}[htbp]
   \centering
      \psfrag{k0}{$\hvec{k}_0$}
      \psfrag{t1}{path 1}
      \psfrag{t2}{path 2}
      \includegraphics[width=6cm]{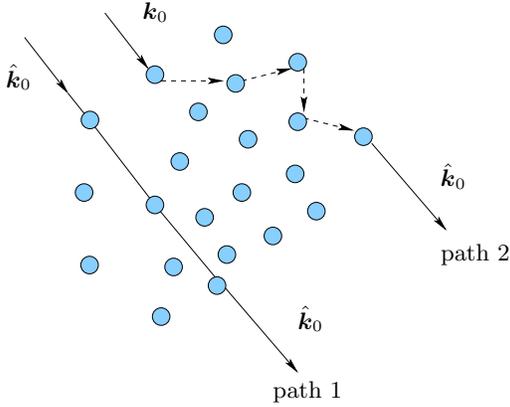}
         \caption{Two different paths which contribute to the
         mass operator $\op{\Sigma}(||\vec{k}_0||,\omega)=n\,\op{C}_{e,o}(\kk_0|\kk_0)$ in the (QC-CPA) approach.
         Only the path of kind 1 are taken into account in  our forward scattering approximation \eref{AppCe}.}
        \label{fig1}
         \end{figure}

To reduce the numerical difficulties in the system of equations
(\ref{PropSigma}-\ref{Obtentionepe}), we add two new
approximations to  the QC-CPA scheme:
\begin{itemize}
\item A far-field approximation:
For an incident plane wave:
\begin{equation}
\E^{i}(\rr)=\E^{i}(\vec{k}_0)\,e^{\rmi\,\vec{k}_0\cdot\rr}\,,
\end{equation}
transverse to the propagation direction $\hvec{k}_0$:
\begin{equation}
\E^{i}(\vec{k}_{0})\cdot\hvec{k}_0=0\,,
\end{equation}
 where $\vec{k}_0=K_e\,\hvec{k}_0$ and
$\hvec{k}_0\cdot\hvec{k}_0=1$,
the scattered far-field, by a
particle within a medium of permittivity $\ep_e(\omega)$, is
 described
by an operator $\op{f}(\hvec{k}|\hvec{k}_0)$ such that:
\begin{equation}
\E^{s}(\rr)=\frac{e^{\rmi\,K_e||\rr||}}{||\rr||}\op{f}(\hvec{k}|\hvec{k}_0)\cdot\E^{i}(\hvec{k}_0)\,.
\label{defkk0}
\end{equation}
 which verifies transversality conditions:
\begin{align}
\op{f}(\hvec{k}|\hvec{k}_0)\cdot\hvec{k}_0 &=0\,,\label{Proptransv1}\\
\hvec{k}\cdot\op{f}(\hvec{k}|\hvec{k}_0) &=0\,.\label{Proptransv2}
\end{align}
 Moreover, the
scattered field in the general case is expressed with the operator
$\op{t}_{e,o}$ by:
\begin{align}
\E^{s}(\rr)=\iintr{1}{2}\op{G}_e^{\infty}(\rr,\rr_1)\cdot\op{t}_{e,o}(\rr_1|\rr_2)\cdot\E^{i}(\rr_2)\,.\no\\\label{Est}
\end{align}
In using the phase perturbation method in equation \eref{Est},
the scattered far-field is obtained in function of the operator
$\op{t}_{e,o}(\vec{k}|\vec{k}_0)$, and in comparing the result
with equation \eref{defkk0}, we obtain the following
relationship: :
\begin{align}
4\pi\,\op{f}(\hvec{k}|\hvec{k}_0)=&(\op{I}-\hvec{k}\hvec{k})\no\\
&\cdot\op{t}_{e,o}(K_e\hvec{k}|K_e\hvec{k}_0)
\cdot(\op{I}-\hvec{k}_0\hvec{k}_0)\,,\label{Chap3lientjf}
\end{align}
Our far-field approximation consist in neglecting the longitudinal
component and the off-shell contribution in the operator
$\op{t}_{e,o}$, and we write:
\begin{align}
\op{t}(K_e\hvec{k}|K_e\hvec{k}_0)&\simeq4\pi\,\op{f}(\hvec{k}|\hvec{k}_0)\,,\label{tfapprox}\\
&=4\pi\,(\op{I}-\hvec{k}\hvec{k})\cdot\op{f}(\hvec{k}|\hvec{k}_0)\cdot(\op{I}-\hvec{k}_0\hvec{k}_0)\,.
\end{align}
where the last equality comes from the properties
(\ref{Proptransv1}-\ref{Proptransv2}).
%Finally, we will neglect
%the term in $1/||\rr||^p$ with $p>1$ in the development of the
%Green function $\op{G}_e^{\infty}(\rr)$, since these terms
%correspond to the non-propagative part of the field.

 \item A forward scattering approximation:
For scatterers large compared to a wavelength, the scattered field
is predominantly in the forward direction (i.e.
$|f(\hvec{k}_0|\hvec{k}_0)|\gg|f(-\hvec{k}_0|\hvec{k}_0)|$). Our
forward approximation consist in keeping only the contribution of
the amplitude of diffusion $f(\hvec{k}|\hvec{k}_0)$ in the
direction of the incident wave $\hvec{k}_0$. We write in using the
hypothesis \eref{tfapprox}:
\begin{align}
\op{t}_{e,o}(K_e\hvec{k}|K_e\hvec{k}_0)&=
4\pi\,\op{f}(\hvec{k}|\hvec{k}_0)\,,\\
&\simeq 4\pi\,\op{f}(\hvec{k}_0|\hvec{k}_0)\,,\label{ForwardApp}\\
&=4\pi\,(\op{I}-\hvec{k}_0\hvec{k}_0)f(K_e,\omega)\,,
\end{align}
where
\begin{equation}
f(K_e,\omega)=\frac{\rmi}{K_e}\,S_1(0)=\frac{\rmi}{K_e}\,S_2(0)\,,
\end{equation}
 with $S_1(0)=S_2(0)$ given by the Mie
theory~\cite{Hulst1,Kerker,Bohren,Ishi2}. It is worth mentioning
that the approximation \eref{ForwardApp} is also valid for small
scatterers (Rayleigh scatterers). In  this case, the scattering
amplitude $f(\hvec{k}|\hvec{k}_0)$ doesn't depend on the direction
of the incident and scattered wave vector  $\hvec{k}$ and
$\hvec{k}_0$, since we have
\begin{equation}
\op{t}_{e,o}(\vec{k}|\vec{k}_0)=t_{e,o}(\omega)\op{I}\,.
\end{equation}
From equation \eref{Chap3lientjf}, we show that:
\begin{equation}
\op{f}(\hvec{k}|\hvec{k}_0)=\op{f}(\hvec{k}_0|\hvec{k}_0)\,.
\end{equation}
and we also obtain the coefficient $f(K_e,\omega)$:
\begin{equation}
4\pi\,f(K_e,\omega)=t_{e,o}(\omega)\,.
\end{equation}
Furthermore, we see from equation \eref{expCkk0}, that to zero
order in density:
\begin{equation}
\op{C}_{e,o}(\vec{k}|\vec{k}_0)=\op{t}_{e,o}(\vec{k}|\vec{k}_0)\,,
\end{equation}
and that the forward approximation \eref{ForwardApp} applied to
the operator $\op{C}_{e,o}(\vec{k}|\vec{k}_0)$ in the low density
limit. We will suppose that the forward approximation is valid
whatever the order in density for the operator
$\op{C}_{e,o}(\vec{k}|\vec{k}_0)$ and we write:
\begin{align}
\op{C}_{e,o}(\kk|\kk_0)&\simeq \op{C}_{e,o}(\kk_0|\kk_0)\,,\\
&\simeq(\op{I}-\hvec{k}_0\hvec{k}_0)\,
C^{\perp}_{e,o}(||\vec{k}_0||,\omega)\,.\label{AppCe}
\end{align}
With this hypothesis, only the path of type 1 in the figure
\eref{fig1} are considered. This approximation also implied that
the operator $\op{C}_{e,o}(\kk|\kk_0)$ is transverse to the
propagation direction $\hvec{k}_0$.
\end{itemize}
From the previous hypothesis and the QC-CPA equations
\eref{expCkk0}, we obtain an equation on
$\op{C}_{e,o}(\kk_0|\kk_0)$:
\begin{align}
\op{C}_{e,o}(\kk_0|\kk_0)=4\pi\,f(K_e,\omega)\op{I}_{\perp}(\hvec{k}_0)\no\,\\
+4\pi\,n\,f(K_e,\omega)\,\op{m}(\vec{k}_0)\cdot\op{C}_{e,o}(\kk_0|\kk_0)\,\label{eqC11-4}
\end{align}
where we have introduce the notation:
\begin{align}
\op{I}_{\perp}(\hvec{k}_0)& =(\op{I}-\hvec{k}_0\hvec{k}_0)\,,\\
\op{m}(\vec{k}_0)& =\intk{1}\,h(\kk_0-\kk_1)\,
\cdot\op{I}_{\perp}(\hvec{k}_0)\cdot\op{G}_{e}^{\infty}(\kk_1)
\cdot\op{I}_{\perp}(\hvec{k}_0)\,.
\end{align}
Then, we have:
\begin{align}
\op{C}_{e,o}(\kk_0|\kk_0)=\left[\op{I}_\perp(\hvec{k}_0)-4\pi\,nf(K_e,\omega)
\op{m}(\vec{k}_0)\right]^{-1}\no\\
\cdot\op{I}_\perp(\hvec{k}_0)\,4\pi\,f(K_e,\omega)\,.\label{eqCe}
\end{align}
In using the classical properties of the Fourier transform, we
write:
\begin{equation}
\op{m}(\vec{k}_0)=\op{I}_\perp(\hvec{k}_0)\cdot\intr{}\e^{-\rmi\vec{k}_0\cdot\rr}\,
h(\rr)\,\op{G}_e^{\infty}(\rr)\cdot\op{I}_\perp(\hvec{k}_0)\,.\label{eqhGe}
\end{equation}
where we have used the translation invariance of the green
function:
$\op{G}_e^{\infty}(\rr-\rr_0)=\op{G}_e^{\infty}(\rr,\rr_0)$.  We
know that the Dyadic Green function has a singularity which can be
separated in introducing the principal value of the Green
function~\cite{Bladel,Kong2001-2,Kong2001-3,Hanson}:
\begin{equation}
\op{G}_e^{\infty}(\rr)=P.V.\op{G}_e(\rr)-\frac{1}{3K_e^2}\,\delta(\rr)\op{I}\,.\label{GreenPV}
\end{equation}
where the principal value is defined by:
\begin{align}
P.V.\intr{0}\op{G}_e^{\infty}(\rr-\rr_0)\cdot\op{\phi}(\rr_0)\no\\
=lim_{a\to 0}\int_{S_a(\rr)}\rmd^3 \rr_0\,
\op{G}_e^{\infty}(\rr-\rr_0)\cdot\op{\phi}(\rr_0)\,,
\end{align}
with $\op{\phi}(\rr_0)$ a test function and  $S_a(\rr)$  a
spherical volume of radius $a$ centered at $\rr$. This principal
value can be easily calculated, and we
obtain~\cite{Kong,Kong2001-3,Vries}:
\begin{align}
P.V.G_e(\rr)=\frac{e^{\rmi K_e||\rr||}}{4\pi||\rr||}\left[\left(
1-\frac{1}{\rmi
K_e||\rr||}-\frac{1}{K_e^2||\rr||^2}\right)\op{I}\no\right.\\
\left.-\left(1-\frac{3}{\rmi K_e||\rr||}-\frac{3}{K_e^2||\rr||^2}
\right)\hvec{r}\hvec{r}\right]\,.
\end{align}
% As previously mentioned, in the far-field approximation we only consider the term in $1/||\rr||$, and equation
% \eref{eqhGe} becomes:
%\begin{align}
%\op{I}_\perp(\hvec{k}_0)\cdot\intk{1}\,h(\kk-\kk_1)\op{G}_e^{\infty}(\vec{k}_1)\cdot\op{I}_\perp(\hvec{k}_0)
%=-\frac{h(0)}{3K_e^2}\op{I}_\perp(\hvec{k}_0)\no\\+\op{I}_\perp(\hvec{k}_0)\cdot\intr{}\e^{-\rmi\vec{k}_0\cdot\rr}\,
%h(\rr)\,\frac{e^{\rmi
%K_e||\rr||}}{4\pi||\rr||}(\op{I}-\hvec{r}\hvec{r})\cdot\op{I}_\perp(\hvec{k}_0)\,,\label{eqhGe-2}
%\end{align}
In using polar coordinate in the integral \eref{eqhGe}:
\begin{equation}
\intr{} \,.=\int_0^{+\infty} r^2 \rmd r\int_{4\pi} \rmd^2 \hvec{r}
\,,
\end{equation}
and the integral on solid angles given in Appendix \ref{appA},
 we obtain the
following result:
\begin{align}
\op{m}(\vec{k}_0)=\left[-\frac{h(0)}{3K_e^2}+\,m(K_e)\right]\op{I}_\perp(\hvec{k}_0)\,.\label{eqhGe-3}
\end{align}
with
\begin{align}
m(K_e)&=\int_0^{+\infty}r\,\rmd r\,p(K_e\,r)\,[g(r)-1]\e^{\rmi K_e
r}\label{Formu2}\,,\\
 p(x)&=\frac{\sin x}{x}-\left(\frac{\sin
x}{x^3} -\frac{\cos
x}{x^2} \right)\,,\no\\
&-\left(\frac{1}{\rmi x}+\frac{1}{x^2}\right)\left(\frac{\sin
x}{x}-3\left[\frac{\sin x}{x^3}-\frac{\cos
x}{x^2}\right]\right)\,.\label{defpx}
\end{align}
where we have assumed that $||\vec{k}_0||=K_e$ since the mass
operator \eref{PropSigma} is evaluated with this value in the
equation \eref{EqKe} to obtain  the effective permittivity.
 As the particle cannot interpenetrate, we have $g(0)=0$ and
then $h(0)=-1$. With equations
(\ref{EqKe},\ref{Propiso},\ref{PropSigma},\ref{eqCe},\ref{eqhGe-3}),
we derive an expression for the effective wave number $K_e$:
\begin{equation}
K_{e}^2=K_{1}^2+\frac{4\pi\,n\,f(K_e,\omega)}{1-4\pi\,n\,f(K_e,\omega)\left(\frac{1}{3K_e^2}+
m(K_e)\right)}\,,\label{Formu1}
\end{equation}
where the scalar $m(K_e)$ is defined  by equations
(\ref{Formu2},\ref{defpx}) and the scalar $f(K_e,\omega)$ is the
forward scattering amplitude:
$\op{f}(\hvec{k}_0|\hvec{k}_0)=(\op{I}-\hvec{k}_0\hvec{k}_0)f(K_e,\omega)$
for a particle of permittivity $\tilde \ep_s=\ep_s-\ep_1+\ep_e$
within a medium of permittivity $\ep_e$ . The relationship between
the effective wave number $K_e$ and the effective permittivity
$\ep_e$ is given by:
\begin{equation}
\ep_e(\omega)=K^2_e/K^2_{vac}\,. \label{Formu3}\end{equation}
 The formula (\ref{Formu2}-\ref{Formu3}) are the main results of
 this paper.
\section{Rayleigh scatterers}
\label{Rayleigh}
 We now show how to recover the low-frequency
limit of the QC-CPA approach.  First, we have to find an
expression for the T matrix for a single scatterer when its size
is small compared to a wavelength ($K_{vac}\,r_s \ll 1$). The T
matrix $\op{t}_{e,o}$ verifies the following equation:
\begin{align}
\op{t}_{e,o}(\rr,\rr_0) =\op{v}_{e,o}(\rr)\,\delta(\rr-\rr_0)+\int
\rmd^3\rr_1\,\,\op{v}_{e,o}(\rr)\no\\ 
\cdot\op{G}^{\infty}_e(\rr,\rr_1) \cdot
\op{t}_{e,o}(\rr_1,\rr_0)\,.\label{Chap4E1transtv}
\end{align}
where the potential is defined by:
\begin{equation}
\op{v}_{e,0}(\rr)\equiv
K_{vide}^2\,(\ep_s-\ep_1)\,\Theta_s(\rr)\,\op{I}\,.\label{Chap4ParDefv1j}
\end{equation}
If we extract the singularity of the Green dyadic function
$\op{G}^{\infty}_e(\rr,\rr_1)$ in using equation
\eref{GreenPV}, we obtain:
\begin{align}
\op{t}_{e,o}(\rr,\rr_0)
&=\op{v}_{dip}(\rr)\,\delta(\rr-\rr_0)+\int
\rmd^3\rr_1\,\,\op{v}_{dip}(\rr)\no\\
&\cdot [P.V.\op{G}^{\infty}_e(\rr,\rr_1)]\cdot
\op{t}_{e,o}(\rr_1,\rr_0)\,,\label{Iterdipole}
\end{align}
where
\begin{align}
\op{v}_{dip}(\rr)&\equiv
\left[1+\frac{\op{v}_{e,o}(\rr)}{3\,K_e^2}
\right]^{-1}\cdot \op{v}_{e,o}(\rr)\,,\\
&=K_{vide}^2\,\alpha_{dip}\,\frac{\Theta_s(\rr)}{\mathtt{v}_s}\,\op{I}\,,\label{potential}
\end{align}
with
\begin{align} \mathtt{v}_s &=\frac{4\pi}{3}r_s^3\\
\alpha_{dip}&=3\,\ep_e\,\frac{\tilde\ep_s-\ep_1}{\tilde\ep_s+2\,\ep_1}\,\mathtt{v}_s\,.
\end{align}
It's easy to recognize that the coefficient $\alpha_{dip}$ is the
polarization factor of a dipole. Hence, the singularity in the
Green dyadic function describes the depolarization factor due to
the induced field in the particles. The relationship between the
singularity of the Green function and the depolarization field
acting on a particle  has been described in numerous
works~\cite{Bladel,Lindell,Kong,Kong2001-3,Tai,Lag2,Vries,Hanson}.
From the meaning of the coefficient $\alpha_{dip}$, we inferred
that equation \eref{Iterdipole} describes the multiple
scattering process by the dipoles inside the particle (where
$\Theta_d(\rr)\neq 0$).
%As the the particle is spherical,
%we can integrate on the solid angle in equation
%\eref{Iterdipole}:
%\begin{align}
%\op{t}_{e,o}(\rr,\rr_0)
%&=\op{v}_{dip}(\rr)\,\delta(\rr-\rr_0)+\int
%r_1^2\rmd r_1\,\,\op{v}_{dip}(r)\no\\
%&\cdot [\int_{4\pi}\rmd^2
%\hvec{r}_1\,P.V.\op{G}^{\infty}_e(\rr,\rr_1)]\cdot
%\op{t}_{e,o}(\rr_1,\rr_0)\,,\label{Iterdipole-2}
%\end{align}

As the particles are small compared to a wavelength, we use a
point scatterer approximation:
\begin{equation}
\frac{\Theta_s(\rr)}{\mathtt{v}_s}\approx \delta(\rr)
\end{equation}
and the potential \eref{potential} becomes:
\begin{equation}
\op{v}_{dip}(\rr)\simeq
K_{vide}^2\,\alpha_{dip}\,\delta(\rr)\,\op{I}\,.\label{pot-2}
\end{equation}
In introducing the approximation \eref{pot-2} in equation
\eref{Iterdipole}, the Dirac distribution provides an analytical
answer for the T matrix of a single particle:
\begin{align}
\op{t}_{e,o}(\rr,\rr_0)&=\delta(\rr-\rr_0)\,\delta(\rr_0)\,\op{t}_{e,o}(\omega)\,,\\
\op{t}_{e,o}(\omega)&=K_{vide}^2\,
\alpha_{dip}\,\left[\op{I}-K_{vide}^2\,\alpha_{dip}\,P.V.\op{G}_e^{\,\infty}(\rr=\vec{0})\right]^{-1}\,.\label{formT}
\end{align}
The principal value of the Green function at the origin can be
evaluated in using a regularization procedure~\cite{Lag2,Vries}:
\begin{equation}
P.V.\,\op{G}_e^{\,\infty}(\rr=\vec{0})=\left[\frac{\Lambda_T}{6\,\pi}+\frac{\rmi\,K_e}{6\,\pi}\right]\op{I}\,,\label{Chap3VPG1en0}
\end{equation}
where the term $\Lambda_T$ is proportional to the inverse of the
real size of the scatterer~\cite{Vries,Lag1}. Finally, the T
matrix for a single particle is:
\begin{align}
\op{t}_{e,o}(\rr_1,\rr_2)& =\delta(\rr_1-\rr_2)\,\delta(\rr_1)\,t_{e,o}(K_e,\omega)\,\op{I}\,,\label{formT2}\\
t_{e,o}(K_e,\omega)&
=\frac{K_{vide}^2\,\alpha_{dip}}{1-K_{vide}^2\,\alpha_{dip}\,
\left(\frac{\Lambda_T}{6\,\pi}+\frac{\rmi\,K_e}{6\pi}\right)}\,,\label{formT3}
\end{align}
It has been shown that the T matrix (\ref{formT2},\ref{formT3})
verifies the optical theorem and can present a resonant behavior
due to the $\Lambda_T$ term~\cite{Vries,Lag1}. The validity of the
optical theorem is an important point, to insure that that the
attenuation of the coherent wave due to scattering is correctly
taken into account in the (QC-CPA) approach. Hence, the
expressions \eref{formT3} must be used rather than the usual T
matrix for a Rayleigh scatterer
${t}_{e,o}(K_e,\omega)=K_{vide}^2\, \alpha_{dip}\,$ which doesn't
verify the optical theorem~\cite{Lag1,Kong2001-3}. Furthermore,
from equations \eref{Chap3lientjf}, we notice that:
\begin{equation}
4\pi\,f(K_e,\omega)=t_{e,o}(K_e,\omega)\,.\label{f=k}
\end{equation}
The small size of the scatterers   allow us also to approximate
the term $m(K_e)$ in equation \eref{Formu1}. In fact, as there
is no long range correlation   in a random medium ($g(r)-1\simeq
0$ for $r\gg r_s$) and as $K_e\,r_s\ll 1$, we can evaluate the
function $p(K_e r)$ in the integral \eref{Formu2} for $K_e r$
close to zero. In using the limit:
\begin{equation}
p(x)=\frac{2}{3}+o(x)\quad x\to 0\,,
\end{equation}
we obtain the following leading term of the real and imaginary
part of  $m(K_e)$:
\begin{align}
m(K_e)=\frac{2}{3}\int_0^{+\infty}r\,\rmd
r\,\,[g(r)-1]\no\\
+\frac{2\rmi\,K_e}{3}\int_0^{+\infty}r^2\,\rmd
r\,[g(r)-1]+\dots\,.
\end{align}
If we keep only these two terms, equation \eref{Formu1}
becomes in using the results (\ref{f=k},\ref{formT3}):
\begin{align}
&\ep_e=\ep_e\no\\
&+\frac{3(\tilde\ep_s-\ep_1)\,\ep_e\,f_{v}}{(\tilde\ep_s-\ep_1)\,(1-f_{v}-\frac{2}{3}(\ep_e^{1/2}\,K_{vac}\,r_s)^3\,
[\frac{w_1}{K_e}+\rmi w_2])+3\,\ep_e}\,,\label{eqray}
\end{align}
with $f_v=n\,\mathtt{v_s}$ the fractional volume occupied by the
particles and $w_1$, $w_2$ defined by:
\begin{align}
4\pi\,n \int_{0}^{+\infty} r\,\rmd r \,[g(r)-1]=w_1-\Lambda_T\,,\label{w1}\\
4\pi\,n \int_{0}^{+\infty} r^2 \rmd r
\,[g(r)-1]=w_2-1\,,\label{w2}
\end{align}
For non resonant Rayleigh scatterers, we can neglect the term
$\Lambda_T\simeq 0$, and also neglect the term
$\frac{2}{3}(\ep_e^{1/2}\,K_{vac}\,r_s)^3\, \frac{w_1}{K_e}$
compare to $1-f_v$. Moreover, we have usually $Re(\ep_e)\gg
Im(\ep_e)$ and equation \eref{eqray} can be simplified into:
\begin{align}
&\ep_e=\ep_1+\frac{3(\ep_s-\ep_1)\,\ep_e\,f_{vol}}{(\ep_s-\ep_1)\,(1-f_{vol})+3\,\ep_e}\no\\
&+\rmi\,\frac{2\,(K_{vide}\,r_d)^3\,(\ep_s-\ep_1)^2\,\ep_e^{5/2}\,f_{vol}}{[(\ep_s
-\ep_1) (1-f_{vol})+3\,\ep_e]^2}\,w_2\,,\label{faiblelam}
\end{align}
where  a Percus-Yevick correlation function for $g(r)$
gives~\cite{Kong,Kong2001-2}:
\begin{equation}
w_2=\frac{(1-f_{v})^4}{(1+2\,f_{v})^2}\,.
\end{equation}
The equation \eref{faiblelam} is the usual low-frequency limit of
the QC-CPA approach obtained by Tsang \emph{et
al.}~\cite{Kong,Kong2001-3}. In particular, we see that in the
static-case ($\omega=0$) the imaginary term in the right hand side
of equation \eref{faiblelam} is null, and if we replace the
effective permittivity $\ep_e$ by $\ep_1$ in the right-hand side
of  equation \eref{faiblelam}, we recover the classical
Maxwell Garnett formula:
\begin{equation}
\ep_e=\ep_1+\frac{3(\ep_s-\ep_1)\,\ep_1\,f_{vol}}
{(\ep_s-\ep_1)\,(1-f_{vol})+3\,\ep_1}\,,\label{MaxGar1}
\end{equation}
which is usually written in the following form:
\begin{equation}
\frac{\ep_e-\ep_1}{\ep_e+2\ep_1}=f_v\,\frac{\ep_s-\ep_1}
{\ep_s+2\ep_1}\,.
\end{equation}
In comparing equation \eref{MaxGar1} and \eref{eqray}, we see
that the scattering process modified the Maxwell Garnett formula
by adding a new term:
\begin{equation}
-\frac{2}{3}(\ep_e^{1/2}\,K_{vac}\,r_s)^3\, [\frac{w_1}{K_e}+\rmi
w_2])\,,
\end{equation}
whose imaginary part describes the attenuation of the coherent
wave, and then, the transfer  to the incoherent part due to the
scattering of the wave.

\section{Keller formula}
\label{Keller} We are now going to show that the  relation
\eref{Formu1} that we have obtained contain also the Keller
formula~\cite{Keller1}. This formula has recently been shown to be
in good agreement with experimental results for particles larger
than a wavelengh~\cite{Hespel,Hespel2}. The Keller formula can be
obtained in considering the QC-CPA approach in the scalar
case~\cite{Kong,Kong2001-3}. The equations are formerly identical
to equation (\ref{PropSigma}-\ref{fourierG1}) where the dyadic
Green function $\op{G}_e^{\infty}$:
\begin{equation}
\op{G}^{\infty}_e(\rr,\rr_0,\omega)=\left[\op{I}+\frac{\nabla\nabla}{K_{e}^2}\right]
\frac{e^{\rmi\,K_{e}\,||\rr-\rr_0||}}{4\pi||\rr||}\,,
\end{equation}
 have to be replaced by the scalar
Green function $G_e^{\infty}$ given by:
\begin{equation}
G^{\infty}_e(\rr,\rr_0,\omega)=
\frac{e^{\rmi\,K_{e}\,||\rr-\rr_0||}}{4\pi||\rr||}\,.
\end{equation}
The first iteration of the scalar version of equation
\eref{expCkk0} gives in using equation \eref{App1}:
\begin{align}
&K_{e}^2=K_{1}^2+4\pi\,n\,f(K_e,\omega)\no\\
&+(4\pi)^2\,n^2\,f^2(K_e,\omega)\int_0^{+\infty}\rmd r\,\frac{\sin
K_e\,r}{K_e}[g(r)-1]\e^{\rmi K_e\,r}+\dots\,.\label{formuKeller}
\end{align}
As was shown by Waterman \emph{et al}, this development is valid
if the following condition is verified:
\begin{equation}
(4\pi)^2\,n\,|f(K_e,\omega)|^2/K_e\ll 1\,.\label{condGeo}
 \end{equation}
In the geometric limit, the scattering cross section $\sigma_s$
for a single particle is in good approximation given by
$\sigma_s\simeq 2\,\pi\,r_s^2$. As the cross section is connected
to scattering amplitude by the relation
$\sigma_s=\frac{8\pi}{3}|f(K_e,\omega)|^2$ and as the maximum
density is $n=1/v_s$, we see that  for particles larger than a
wavelength the condition \eref{condGeo} is satisfied:
\begin{equation}
(4\pi)^2\,n\,|f(K_e,\omega)|^2/K_e \simeq 1/K_e\,r_s\ll 1\,.
\end{equation}
The equation \eref{formuKeller}, which has been derived by
Keller~\cite{Keller1}, has proven to be in good agreement with
experiments for particles larger than a wavelength. If we now use
a Taylor development in equation \eref{Formu1}, we obtain
\begin{align}
&K_{e}^2=K_{1}^2+(4\pi)^2\,n\,f(K_e,\omega)\no\\
&+(4\pi)^2\,n^2\,f^2(K_e,\omega)\left[\frac{1}{3\,K_e^2}+m(K_e)\right]+\dots\,.\label{KellerVect}
\end{align}
This development is valid if the condition \eref{condGeo} is
satisfied. In the geometric limits, we can approximate the
function $p(K_e\,r)$ in the definition \eref{Formu2} of $m(K_e)$,
since for $K_e\,r\gg 1$ we have from equation \eref{defpx}:
\begin{equation}
p(x)\simeq \frac{sin x}{x}\,, \quad x\gg 1,
\end{equation}
The relation \eref{KellerVect} becomes:
\begin{align}
&K_{e}^2=K_{1}^2+4\pi\,n\,f(K_e,\omega)\no\\
&+(4\pi)^2\,n^2\,f^2(K_e,\omega)\left[\frac{1}{3\,K_e^2}+\int_0^{+\infty}\rmd
r\,\frac{\sin K_e\,r}{K_e}[g(r)-1]\e^{\rmi
K_e\,r}\right]\,.\label{KellerVect2}
\end{align}
We see that equation \eref{KellerVect2} differ from the
equation \eref{formuKeller}, only by the factor $1/3K_e^2$, which
is due to singularity of the vectorial Green function and
consequently cannot be derived from the scalar theory developed by
Keller. We also remarks that that to solve numerically  our new
equation \eref{Formu1}, we can use the same procedure that is used
to solve the original Keller formula \eref{formuKeller} with the
Muller theory~\cite{Hespel2}. Consequently, we have  derived a
numerical tractable approximation to the (QC-CPA) scheme.
%For large particle, we can neglect the $1/K_e^2$ since for example
%in hte case of hard sphere correlation model where
%$g(r)=\theta(r)$, we have:
%\begin{align}
%\int_0^{+\infty}\rmd r\,\frac{\sin K_e\,r}{K_e}[g(r)-1]\e^{\rmi
%K_e\,r} &=\frac{1}{2iK_e^2}\left[\frac{(\e^{2\rmi
%K_e r_s}-1)}{2\rmi}-K_e r_s\right]\,,\\
%&\simeq -\frac{\e^{\rmi K_e\,r_s}}{4 K_e^2}
%\end{align}
%which is larger than the $1/K_e^2$ in equation for $K_e r_s\gg
%1$.
\section{Conclusion}
The intent of this paper has been to establish a new formula for
the effective dielectric constant which characterize the coherent
part of  an electromagnetic wave propagating in a random medium.
The starting point of our theory has been the quasicrystalline
coherent potential approximation which takes into account the
correlation between the particles. As the  numerical calculation
of the effective permittivity  is still a difficult task under the
(QC-CPA) approach, we have added a far-field and a forward
scattering approximations to (QC-CPA) scheme. In the low frequency
limit, equation is identical  with  the usual result obtained
under the (QC-CPA) scheme, and in the high frequency limit  the
expression include the generalization, in the vectorial case, of
the result obtained by Keller. Further study is necessary to
assess the limitation of this approach on the intermediate
frequency regime.

\appendix
\section{Appendixes}
\label{appA}
\begin{align}
\int_{4\pi}\rmd^2\hvec{r}\,e^{-\rmi\,\vec{k}_0\cdot\hvec{r}||\rr||}=4\pi\,\frac{\sin
||\vec{k}_0||\,||\rr||}{||\vec{k}_0||\,||\rr||}\,,\label{App1}
\end{align}

\begin{align}
&\int_{4\pi}\rmd^2\hvec{r}\,e^{-\rmi\,\vec{k}_0\cdot\hvec{r}||\rr||}\hvec{r}\hvec{r}=4\pi\left[\frac{\sin
||\vec{k}_0||\,||\rr||}{||\vec{k}_0||^3\,||\rr||^3}\right.\no\\&\left.-\frac{\cos
||\vec{k}_0||\,||\rr||}{||\vec{k}_0||^2\,||\rr||^2}\right](\op{I}-\hvec{k}_0\hvec{k}_0)
+4\pi\left[\frac{\sin
||\vec{k}_0||\,||\rr||}{||\vec{k}_0||\,||\rr||}\right.\no\\
&\left.+2\frac{\cos
||\vec{k}_0||\,||\rr||}{||\vec{k}_0||^2\,||\rr||^2}-2\frac{\sin
||\vec{k}_0||\,||\rr||}{||\vec{k}_0||^3\,||\rr||^3}\right]\hvec{k}_0\hvec{k}_0\,,\label{App2}
\end{align}

\newpage %Just because of unusual number of tables stacked at end
%\bibliography{bibl}
\bibliography{Livres_Bib,Articles_Bib}
\end{document}